\documentclass[conference]{IEEEtran}

\usepackage{amsmath,amsthm}
\usepackage[cmintegrals]{newtxmath}

    \theoremstyle{definition}
	\newtheorem{example}{Example}[]

\usepackage[colorlinks=true,bookmarks=false,citecolor=blue,urlcolor=blue]{hyperref} 
\usepackage{multirow}
\usepackage[dvipsnames]{xcolor}
\usepackage{mathtools} 
\usepackage{blkarray, bigstrut} %
\usepackage{adjustbox}
\usepackage{cite}
\usepackage{setspace}

\usepackage{setspace}
\usepackage{color, colortbl}

\colorlet{pruneclr}{green!70!black}%

\DeclareMathAlphabet{\pazocal}{OMS}{zplm}{m}{n}
\SetMathAlphabet\pazocal{bold}{OMS}{zplm}{bx}{n}

\def\lf{\left\lfloor}
\def\rf{\right\rfloor}
\def\lc{\left\lceil}
\def\rc{\right\rceil}

\def\boldA{\boldsymbol{A}}
\def\boldb{\boldsymbol{b}}
\def\boldB{\boldsymbol{B}}
\def\boldc{\boldsymbol{c}}
\def\boldC{\boldsymbol{C}}

\def\calA{\pazocal{A}}

\def\n{N}

\def\emax{E^\bullet}

\def\Rs{R_{\text{s}}}

\def\hi{h_{\text{i}}}
\def\hs{h}
\def\wi{w_{\text{i}}}
\def\wf{w_{\text{f}}}

\usepackage{tikz}
    \usetikzlibrary{positioning}
    \usetikzlibrary{calc}
    \usetikzlibrary{shapes}
    \usetikzlibrary{patterns}
    
\usepackage{pgfplots,pgfplotstable}
    \pgfplotsset{compat=1.15}
    \usepgfplotslibrary{fillbetween}
\usetikzlibrary{positioning, decorations.pathreplacing, calc, intersections, pgfplots.groupplots}

\usepackage{xstring}

\def\lf{\left\lfloor}
\def\rf{\right\rfloor}

\def\emax{E_{\max}}

\newcounter{mylabelcounter}

\makeatletter
\newcommand{\labelText}[2]{%
#1\refstepcounter{mylabelcounter}%
\immediate\write\@auxout{%
  \string\newlabel{#2}{{1}{\thepage}{{\unexpanded{#1}}}{mylabelcounter.\number\value{mylabelcounter}}{}}%
}%
}
\makeatother

\begin{document}

\title{Band-ESS: Streaming Enumerative Coding with Applications to Probabilistic Shaping}
\author{\IEEEauthorblockN{Yunus Can G\"{u}ltekin, Frans M. J. Willems, Alex Alvarado}
\IEEEauthorblockA{Information and Communication Theory Lab, Eindhoven University of Technology, The Netherlands\\
Email: y.c.g.gultekin@tue.nl}
}

\maketitle

\begin{abstract}
Probabilistic amplitude shaping (PAS) is on track to become the de facto coded modulation standard for communication systems aiming to operate close to channel capacity at high transmission rates.
The essential component of PAS that breeds this widespread interest is the amplitude shaping block, through which the channel input distribution is controlled.
This block is responsible for converting bit strings into amplitude sequences with certain properties, e.g., fixed composition, limited energy, limited energy variation, etc.
Recently, band-trellis enumerative sphere shaping (B-ESS) was introduced as an amplitude shaping technique that achieves limited energy variations which is useful in optical communication scenarios.
B-ESS operates based on a trellis diagram in which sequences with high energy variations are pruned.
In this work, we study the implementation of B-ESS.
We first show that thanks to the trellis structure obtained by this pruning, B-ESS can be implemented with very low storage complexity.
The trellis computation is shown to be reduced to a set of recursive multiplications with a scalar factor.
Then we show that this scalar factor can be adjusted such that the trellis computation is further simplified and realized with only binary shifts.
This \emph{shift-based} B-ESS (1) can be implemented for arbitrarily long blocklengths without incurring an increase in complexity, and (2) can operate in a streaming mode similar to convolutional coding.
\end{abstract}

\begin{IEEEkeywords}
Probabilistic Amplitude Shaping, Enumerative Coding, Streaming Encoder.
\end{IEEEkeywords}

\IEEEpeerreviewmaketitle

\section{Introduction}
\begin{figure}[t]
    \centering
\includegraphics[width=\columnwidth]{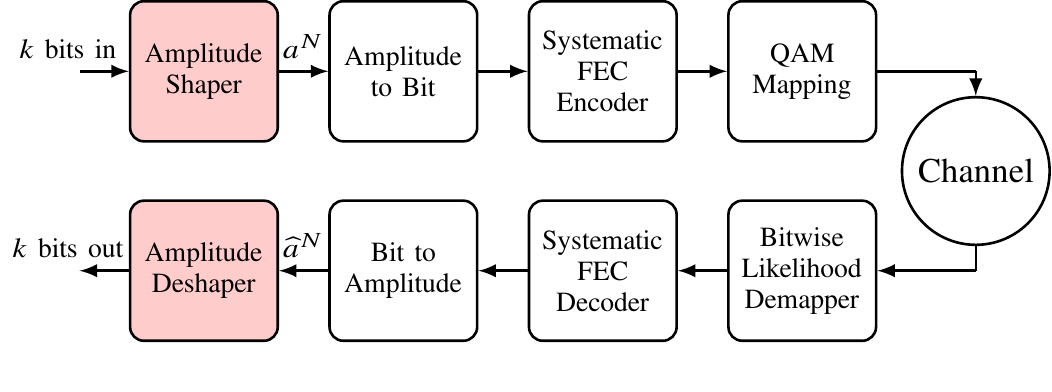}
    \caption{PAS block diagram. Red blocks are the focus of this paper.}
    \label{fig:pas}
\end{figure}

Probabilistic amplitude shaping (PAS) is a coded modulation strategy that combines constellation shaping and forward error correction (FEC)~\cite{BochererSS2015_ProbAmpShap}, and achieves the capacity of the additive white Gaussian noise (AWGN) channel~\cite{Bocherer2014_ProbSigShapForBMD}.
The shaping block generates the amplitudes of the channel inputs while a systematic FEC encoder determines their signs as shown in Fig.~\ref{fig:pas}.
PAS has attracted considerable attention and been shown to increase the achievable rates for a variety of channels, e.g., wireless~\cite{Schulte2019_SMDM,GultekinHKW2019_ESSforShortWlessComm}, optical fiber~\cite{Buchali2016_RateAdaptReachIncrease_letter,Amari2019_IntroducingESSoptics_letter}, free-space optical~\cite{Elzanaty2020_PCSimddFSO}, underwater optical~\cite{Peng2020_PSNLunderwaterVLC}, etc.

Enumerative sphere shaping (ESS) is an amplitude shaping algorithm based on enumerative coding~\cite{GultekinHKW2019_ESSforShortWlessComm}.
ESS converts a binary string into an $\n$-amplitude sequence that has energy no greater than a certain threshold, i.e., sequences are located within an $\n$-sphere.
This approach---sphere shaping---is optimum for the AWGN channel asymptotically for large $\n$ and large constellation cardinality $M$ in the sense that the resulting input distribution converges to a Gaussian distribution.
For finite $M$, the distribution converges to Maxwell-Boltzmann (MB) for large enough $\n$, which has been shown to virtually maximize achievable rates for $M$-amplitude-shift keying ($M$-ASK) input alphabets~\cite{BochererSS2015_ProbAmpShap}.
Thus, ESS has been considered for wireless~\cite{GultekinHKW2019_ESSforShortWlessComm} and long-haul optical channels~\cite{Amari2019_ESSreachincrease_letter,Goossens2019_FirstExperimentESS_letter}, and shown to provide more than 1 dB signal-to-noise ratio (SNR) gains and more than 30\% optical reach increase, resp.

ESS is a special case of the general class of enumerative coding algorithms (e.g.,~\cite{Cover1973_EnumSourceEncode},~\cite{Schalkwijk1972_SourceCoding}) focusing on spherical signal spaces and Gaussian signaling.
However, via proper modifications in the ESS algorithm, other types of signal space constructions and input distributions can be obtained.
For instance, we have recently introduced kurtosis-limited ESS (K-ESS)~\cite{gultekin_kess_arxiv_letter} which introduces a constraint on the maximum fourth power of the norm of amplitude sequences (i.e., kurtosis) in addition to the sphere constraint on their squared norm (i.e., energy) and hence, achieves Gaussian-like input distributions with smaller kurtosis.
Such distributions provide higher achievable rates than MB distributions for short-haul optical links~\cite{Sillekens2018_nonlinearitytailoredPS_letter,Cho2016_ShapingNtolerance} and thus, K-ESS has been shown to provide SNR gains and reach increases over ESS. 
We have also recently proposed \emph{band-trellis ESS} (B-ESS)~\cite{gultekin_bess_ofc_arxiv}.
B-ESS modifies the ESS algorithm such that shaped channel inputs with limited energy variations are generated.
Transmitted signals with high energy variations generate a larger nonlinear interference during propagation over fiber~\cite{Geller2016_ShapingNLphaseNoise,Cho2021_ShapingLightwaves_letter}.
Consequently, we demonstrated that B-ESS provides significant SNR gains over ESS and K-ESS, and up to 10\% rate increase.

Increasing the shaping blocklength $N$ has been shown to improve the performance for linear channels, and also for the nonlinear fiber channel when there is a carrier phase recovery algorithm implemented~\cite{Civelli2020_interplayPSandCPR}.
However, for large values of $\n$, the storage complexity of all the previous algorithms scales poorly~\cite[Sec. 6.2]{Gultekin2019Arxiv_Comparison}.
Moreover, most shaping algorithms (e.g., the ones based on ESS or on constant composition distribution matching~\cite{SchulteB2016_CCDM}) are inherently serial and block-based~\cite[Sec. 6.1]{Gultekin2019Arxiv_Comparison}.
In this paper, as we hinted in~\cite{gultekin_bess_ofc_arxiv}, we propose to diverge from this block-based high-storage-complexity shaping paradigm.
We introduce a \textit{streaming} B-ESS implementation, similar to convolutional encoding.
This implementation has two main advantages: (1) the shaping blocklength $\n$ can be increased arbitrarily without an increase in storage complexity, and (2) most of the computations required to implement the shaping stage reduce to binary shifts.
Moreover, as B-ESS is a generalization of ESS, this low-complexity shift-based streaming implementation applies to ESS as well.

\section{Limiting Signal Energy Variations via ESS}
\subsection{Enumerative Sphere Shaping}
\begin{figure}[t]
    \centering
    \includegraphics[width=0.9\columnwidth]{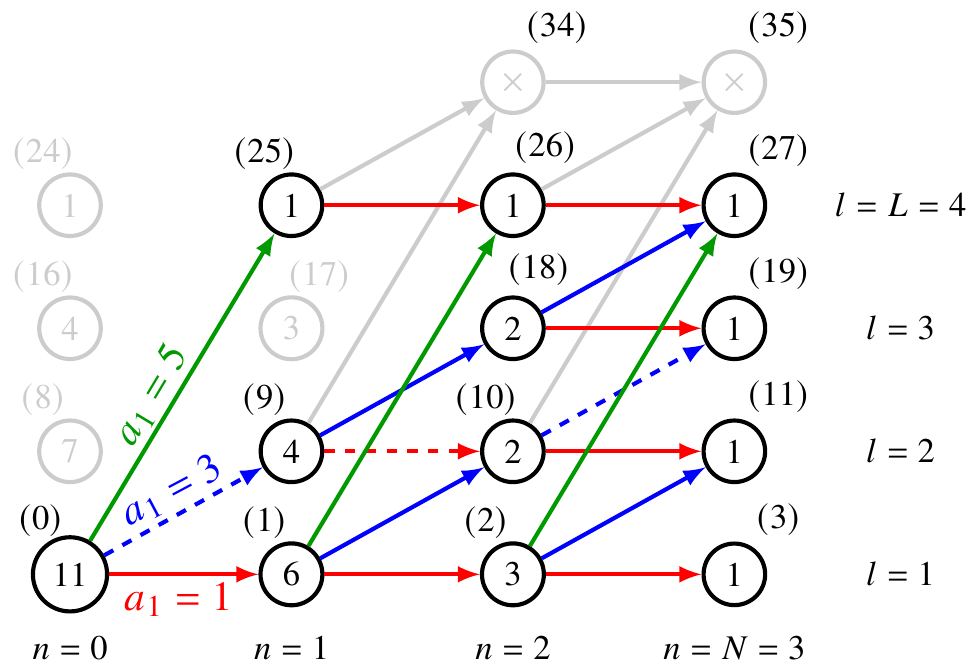}
    \caption{ESS trellis for $N=3$, $\calA=\{1, 3, 5\}$, and $\emax=8(L-1)+\n=27$.
    The sequences that have energy larger than $27$ (e.g., $(5, 3, 1)$) do not satisfy the sphere constraint, and thus, they are not included in the trellis.}
    \label{fig:ess_trellis}
\end{figure}

ESS indexes the sequences in a sphere shaping set
\begin{equation}
\calA^\bullet = \left\{ a^\n = (a_1, a_2, \cdots, a_\n): \sum_{n=1}^{\n} a_n^2 \leq \emax \right\} ,\label{eq:sphereset}
\end{equation} 
where $a_i\in\calA$, and $\calA = \{1, 3,\dotsc, M-1\}$ is the amplitude alphabet of $M$-ASK~\cite{GultekinHKW2019_ESSforShortWlessComm}.\footnote{Typically, the inputs of a wireless communication channel are drawn from an $M^2$-ary quadrature amplitude modulation ($M^2$-QAM) alphabet, which is the Cartesian product of an $M$-ASK alphabet with itself.
Dual-polarized inputs of optical communication channels, however, are often drawn from the Cartesian product of an $M^2$-QAM alphabet with itself.}
For mapping $k$-bit strings to $a^\n\in\calA^\bullet$ invertibly, a trellis is created as shown in Fig.~\ref{fig:ess_trellis}.
Here, $L$ is the number of nodes in the final column, i.e., the {\it height} of the trellis, while the width of the trellis is $N+1$.
The maximum energy $\emax$ can then be written~\cite{GultekinHKW2019_ESSforShortWlessComm} as $\emax = 8(L-1)+\n$ for $M$-ASK alphabets. 
The nodes in the trellis are specified by the pair $(n,l)$ with $n=0, 1,\dotsc, \n$ and $l=1, 2, \dotsc, L$.

In Fig.~\ref{fig:ess_trellis}, the nodes $(n,l)$ are labeled with $(e)$, the energy level $e = 8(l-1)+n$ that they represent, which is the the accumulated energy of the sequences for their first $n$ amplitudes, $\sum_{i=1}^n a_i^2$.
Branches that arrive at a node $(n,l)$ in the $n^{\text{th}}$ column represent $a_n$.
Each $\n$-sequence is, therefore, represented by an $\n$-branch path, starting from node $(0,1)$ and arriving at a node $(N,l)$ in the final, i.e., rightmost, column.
As an example, the dashed path in Fig.~\ref{fig:ess_trellis} represents the sequence $(3, 1, 3)$ with energy $e=8(3-1)+3=19$.

The values written inside each node $(n,l)$ give the number of ways to arrive at the final column, starting from $(n,l)$.
We denote the column vector that stores the values in the $n^{\text{th}}$ column of the trellis by $\boldsymbol{c}_n = [c_{L,n}\hspace{1.5mm} c_{L-1,n}\hspace{1.5mm} \dotsc\hspace{1.5mm} c_{1,n}]^T$ where $c_{l,n}$ is the value inside the node $(n,l)$. 
These values can then be stored in the form of a matrix $\boldC = [\boldc_0\hspace{0.1cm}  \boldc_1\hspace{0.1cm}  \dotsc \boldc_N]$, i.e.,
\begin{equation}
\boldC =
\scalebox{0.9}{
    $\begin{bmatrix}
     c_{L,0} & \cdots & c_{L,N} \\
     \vdots & \ddots & \vdots \\
     c_{1,0} & \cdots & c_{1,N} \\
    \end{bmatrix}$}
    \stackrel{\scalebox{0.80}{\mbox{(Fig.~\ref{fig:ess_trellis}})}}{=}
    \scalebox{0.9}{$\begin{bmatrix}
    \textcolor{orange}{1} & \textcolor{green!70!black}{1} & \textcolor{green!70!black}{1} & \textcolor{green!70!black}{1} \\
    \textcolor{orange}{4} & \textcolor{orange}{3} & \textcolor{green!70!black}{2} & \textcolor{green!70!black}{1} \\
    \textcolor{orange}{7} & \textcolor{green!70!black}{4} & \textcolor{green!70!black}{2} & \textcolor{green!70!black}{1} \\
    \textcolor{green!70!black}{11} & \textcolor{green!70!black}{6} & \textcolor{green!70!black}{3} & \textcolor{green!70!black}{1} \\
    \end{bmatrix}$}.
    \label{eq:exampletrelmat}
\end{equation}
Note that $\boldC$ has ``reverse'' row indexing to indicate that upper rows correspond to higher energy levels as in Fig.~\ref{fig:ess_trellis}.
Moreover, the values inside nodes that correspond to energy levels that are not possible (written in light gray in Fig.~\ref{fig:ess_trellis}) are written in orange in \eqref{eq:exampletrelmat} and not used for shaping, while the others are green.
For instance, the node $(n=1,l=3)$ requires an amplitude with energy $17$ and thus, it is unconnected in Fig.~\ref{fig:ess_trellis}.

The matrix $\boldC$ can be computed recursively via
\begin{equation}
\boldc_n = \widetilde{\boldsymbol{A}} \boldc_{n+1}, \label{eq:ess_transformation}
\end{equation}
for $n = N-1, N-2,\cdots, 0$, where the initial condition is $\boldc_N = [1\hspace{1.5mm} 1\hspace{1.5mm} \dotsc\hspace{1.5mm} 1]$.
Here, we call the $L$-by-$L$ matrix 
\begin{equation}
\widetilde{\boldsymbol{A}} =
\scalebox{0.9}{$\begin{bmatrix}
     a_{L,L} & \cdots & a_{L,1} \\
     \vdots & \ddots & \vdots \\
     a_{1,L} & \cdots & a_{1,1} \\
    \end{bmatrix}$}
    \stackrel{\scalebox{0.80}{\mbox{(Fig.~\ref{fig:ess_trellis}})}}{=}
\scalebox{0.9}{$\begin{bmatrix}
1&0&0&0 \\
1&1&0&0 \\
0&1&1&0 \\
1&0&1&1 \\
\end{bmatrix}$},
\label{eq:ess_adjacency}
\end{equation}
the {\it ESS adjacency matrix}.
In $\widetilde{\boldsymbol{A}}$, $a_{ij}=1$ if there is a connection between the nodes $(n,i)$ and $(n+1,j)$ (i.e., $\widetilde{\boldsymbol{A}}$ has reverse indexing for both rows and columns).
The structure of this adjacency matrix depends on $\calA$ and $L$.

Based on $\boldC$, ESS realizes a mapping from $k$-bit strings to $a^\n\in\calA^\bullet$~\cite{GultekinHKW2019_ESSforShortWlessComm}.
The input length of ESS and the corresponding shaping rate are defined as $k = \lf \log_2 c_{1,0}\rf$ (bits) and $\Rs = k/N$ (bits per amplitude, b/amp), respectively.

\subsection{Approximate Enumerative Sphere Shaping}\label{ssec:bp_ess}
When $\boldC$ is computed using \eqref{eq:ess_transformation}, its elements satisfy
\begin{equation}
    c_{i,n} \leq \sum_{j=1}^{L} a_{ij}c_{j,n+1}, \label{eq:ess_summation}
\end{equation}
with strict equality.
In~\cite{GultekinWHS2018_ApproxEnumerative}, we showed that the invertibility of ESS is guaranteed for any $\boldC$ that satisfies~\eqref{eq:ess_summation}.
For the same set of parameters $N$, $\calA$, and $L$, the only difference between the {\it exact} $\boldC$ that satisfies~\eqref{eq:ess_summation} with equality and an {\it approximate} $\boldC$ that does not is that some sequences inside the sphere are lost in the approximate case.
This is in the sense that the ESS algorithm operating on an approximate $\boldC$ cannot output some sequences from the sphere, which causes a rate loss. 
This creates a trade-off between performance (via rate) and complexity (via required precision and storage) as follows.
After its computation, the binary representation of $c_{l,n}$ is rounded down to $n_{\alpha}$ binary digits (denoted by $\lf c_{l,n} \rf_{n_{\alpha}}$).
Then $c_{l,n}$ is approximated as
\begin{equation}
    c_{l,n} \approx \lf c_{l,n} \rf_{n_{\alpha}} = \alpha 2^\beta, \label{eq:approxESS}
\end{equation}
where $\alpha$ is an $n_{\alpha}$-binary-digit-long mantissa, and $\beta$ is an $n_{\beta}$-binary-digit-long exponent.
More precisely, this approximation rounds the least significant $n_{\beta}$ binary digits of $c_{l,n}$ down to zero, while keeping the most significant $n_{\alpha}$ binary digits unchanged.
Then the number $c_{l,n}$ can approximately be stored in the form $(\alpha,\beta)$ using $(n_{\alpha}+n_{\beta})$~bits.
Note that this is a $(n_{\alpha}+n_{\beta})$-bit base-2 floating-point number representation format which we call \emph{bounded precision (BP)} in which only nonnegative integers are represented.
The exponent length is
\begin{equation}
    n_\beta = \lc \log_2(k+1-n_\alpha) \rc \quad \mbox{(binary digits).} \label{eq:approx_ess_exponent}
\end{equation}
The resulting rate loss is upper-bounded by $\log_2\left(2^{n_\alpha-1}\right) - \log_2\left(2^{n_\alpha-1}-1\right)$ b/amp~\cite{GultekinWHS2018_ApproxEnumerative}.
By proper selection of $n_{\alpha}$, the required storage for $\boldC$ can be decreased significantly with a negligible rate loss.

\subsection{Band-trellis: Altering the Trellis to Limit Energy Variations}
\begin{figure*}[t]
    \hspace{-0.3cm}
        \includegraphics[width=0.65\textwidth]{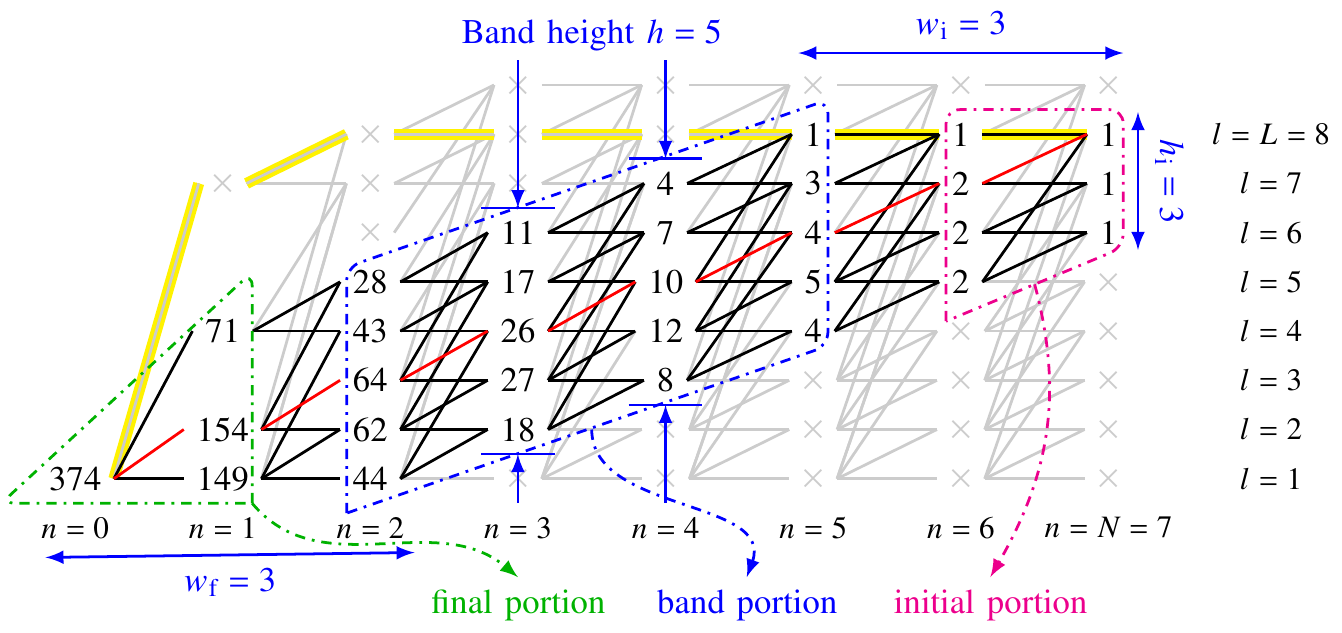}
    \hspace{-0.55cm}
    \resizebox{0.38\textwidth}{!}{\def\sx{1.5}
\def\sy{0.5}
\begin{tikzpicture}[line width=0.8pt]
\node [] (a0) at (0,0) {$\begin{aligned}
\boldC =
\begin{bmatrix}
    0 & 0 & 0 & 0 & 0 & \textcolor{blue}{1} & \textcolor{magenta}{1} & \textcolor{magenta}{1} \\
    0 & 0 &0 & 0 &\textcolor{blue}{4} & \textcolor{blue}{3}&\textcolor{magenta}{2} & \textcolor{magenta}{1}\\
    0&0&0&\textcolor{blue}{11}&\textcolor{blue}{7}&\textcolor{blue}{4}&\textcolor{magenta}{2}&\textcolor{magenta}{1}\\
    0&0&\textcolor{blue}{28}&\textcolor{blue}{17}&\textcolor{blue}{10}&\textcolor{blue}{5}&\textcolor{magenta}{2}&0\\
    0&\textcolor{green!70!black}{71}&\textcolor{blue}{43}&\textcolor{blue}{26}&\textcolor{blue}{12}&\textcolor{blue}{4}&0&0\\
    0&0&\textcolor{blue}{64}&\textcolor{blue}{27}&\textcolor{blue}{8}&0&0&0\\
    0&\textcolor{green!70!black}{154}&\textcolor{blue}{62}&\textcolor{blue}{18}&0&0&0&0\\
    \textcolor{green!70!black}{374}&\textcolor{green!70!black}{149}&\textcolor{blue}{44}&0&0&0&0&0\\
    \end{bmatrix}
    \label{eq:bandtrellismatrix}
    \end{aligned}$
    \hspace{0.15cm}
    (\labelText{11}{label:bandtrellismatrix})
    };
\draw [latex-latex,color=blue] (1.8*\sx,3.5*\sy) -- (1.8*\sx,0.85*\sy) node [pos=0.5, right] {$h_\text{i} = 3$};
\draw [latex-latex,color=blue] (0.7*\sx,4*\sy) -- (1.65*\sx,4*\sy) node [pos=0.5, above] {$w_\text{i} = 3$};
\draw [-, color=blue, rounded corners, dashdotted] (-0.7*\sx,-3.75*\sy) -- (0.95*\sx,-0.55*\sy) -- (0.95*\sx,3.85*\sy) -- (-0.7*\sx,0.7*\sy) -- (-0.7*\sx,-3.7*\sy) {};
\draw[-latex,blue,dashdotted] (-0.3*\sx,-2.95*\sy) to [out=-105,in=100] (0.2*\sx,-4.75*\sy);
\draw [draw=none, blue] (0*\sx,-4.75*\sy) -- (0*\sx,-4.75*\sy) node [below] {band portion};
\draw[-latex,blue,dashdotted] (0.1*\sx,2.49*\sy) to [out=80,in=-40] (-0.175*\sx,4.105*\sy);
\draw [draw=none, blue] (-0.175*\sx,4.105*\sy) -- (-0.175*\sx,4.105*\sy) node [above] {$\boldB$};
\draw [-, color=magenta, rounded corners, dashdotted] (1.05*\sx,-0.15*\sy) -- (1.05*\sx,3.65*\sy) -- (1.65*\sx,3.65*\sy) -- (1.65*\sx,1*\sy) -- (1.05*\sx,-0.15*\sy) {};
\draw[-latex,magenta,dashdotted]  (1.05*\sx,-0.15*\sy) to [out=-110,in=70] (1*\sx,-4.75*\sy);
\draw [draw=none, magenta] (1.4*\sx,-4.75*\sy) -- (1.4*\sx,-4.75*\sy) node [below] {initial portion};
\draw [-, color=green!70!black, rounded corners, dashdotted] (-0.8*\sx,-3.45*\sy) -- (-0.8*\sx,1.25*\sy)  -- (-2.15*\sx,-3.45*\sy) -- (-0.8*\sx,-3.45*\sy)  {};
\draw[-latex,green!70!black,dashdotted] (-1.45*\sx,-3.45*\sy) to [out=-120,in=90] (-1.4*\sx,-4.75*\sy);
\draw [draw=none, green!70!black] (-1.4*\sx,-4.75*\sy) -- (-1.4*\sx,-4.75*\sy) node [below] {final portion};
\node [align=center,draw=none] (a0) at (0,-3.15) {};
\end{tikzpicture}}
    \caption{{\bf Left:} Band-trellis for $N=7$, $\calA=\{1, 3, 5, 7\}$, $\emax=63$, $\hi=\wi=3$, and $s=1$. {\bf Right:} The corresponding $\boldC$.} 
    \label{fig:band_trellis}
\end{figure*}
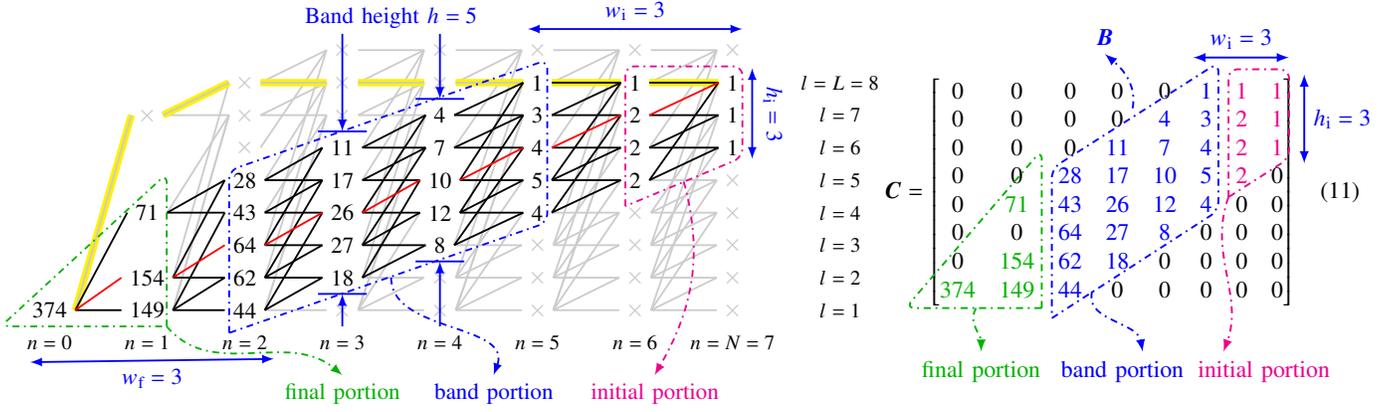

To discard sequences with large energy variations, we consider a band-like portion of the amplitude trellis as shown in Fig.~\ref{fig:band_trellis} (left).
Unlike Fig.~\ref{fig:ess_trellis}, only the values $c_{l,n}$ are included in Fig.~\ref{fig:band_trellis} (left), and the nodes are not labeled with their corresponding energy levels for the sake of simplicity.

The motivation behind the modification in Fig.~\ref{fig:band_trellis} (left) is that the sequences represented in this band will have smaller energy variance $\text{Var}(A^2)$ than the others.
As an example, consider the sequence $(7,3,1,1,1,1,1)$ with energy $e=8(l-1)+n=63$, which is highlighted by yellow in Fig.~\ref{fig:band_trellis} (left).
This sequence is in the complete trellis, but not in the band.
Also, consider the sequence $(3,3,3,3,3,3,3)$ which is drawn with red, belongs to the band, and also has energy $63$.
The first sequence has an energy variance $\text{Var}(A^2)=274.3$, while the second has $0$.
Thus, we expect the sequences in the band to have a lower average energy variance than the ones in the complete sphere. 
The band-trellis can then be seen as a {\it double pruning} of sequences of uniform signaling which has an $N$-cubical signal set.
The first is due to the sphere constraint imposed by $\emax$ which leads to Fig.~\ref{fig:ess_trellis}.
The second is due to the newly introduced band structure which then leads to Fig.~\ref{fig:band_trellis} (left).

We specify the band-trellis with three values (in addition to $N, \calA,$ and $L$): 
(i) the number of active nodes $h_{\text{i}}$ (initial height), (ii) $w_{\text{i}}$ (initial width), and (iii) the slope $s$ of the band (in nodes per column).
For the example in Fig.~\ref{fig:band_trellis}, $h_{\text{i}}=w_{\text{i}}=3$ (in the top-right portion of the trellis) and $s=1$.
From these parameters, two other parameters can be derived: $w_{\text{f}} = N+1-(L-h_{\text{i}})/s$ (final width) on the bottom-left portion of the trellis, and $\hs = \hi + s(w_{\text{i}}-1)$ (the height of the band).
The parameter $\hs$ will be used when computing the trellis via matrix multiplications in the next section.
The parameter $w_{\text{f}}$, on the other hand, can be used to check the validity of the design choices since $w_{\text{f}}\in\mathbb{N}$.
Finally, since the values in the trellis is computed from right to left, we call the rightmost $\wi-1$ columns of the trellis the {\it initial} portion (dash-dotted magenta trapezium in Fig.~\ref{fig:band_trellis}), the leftmost $\wf-1$ columns the {\it final} portion (dash-dotted green triangle in Fig.~\ref{fig:band_trellis}), and the middle $N-\wi-\wf+3$ columns the {\it band} portion (dash-dotted blue parallelogram in Fig.~\ref{fig:band_trellis}).

We call ESS operating on band-trellises {\it band-ESS} (B-ESS).
We note that for increasing $\hi$ and $\wi$, the band-trellis approaches a complete trellis, and hence, B-ESS approaches ESS.
In this sense, we say that B-ESS generalizes ESS.
We demonstrated in~\cite{gultekin_bess_ofc_arxiv} that B-ESS creates a trade-off between the energy efficiency $E[A^2]$ which is important for, e.g., the AWGN channel, and energy variations $\text{Var}(A^2)$ which is important for, e.g., the nonlinear fiber channel.
Thanks to this trade-off, B-ESS provided higher SNRs than uniform signaling and ESS, and an up to 10\% achievable rate increase.

\section{Streaming Shaping Encoders with B-ESS}
The values in a band-trellis can again be stored in the form of a matrix $\boldC$.
For the example in Fig.~\ref{fig:band_trellis} (left), $\boldC$ is given in~(\nameref{label:bandtrellismatrix}), see Fig.~\ref{fig:band_trellis} (right).
The entries of $\boldC$ that correspond to the initial, band, and final portions of the band trellis are color-coded in~(\nameref{label:bandtrellismatrix}) similar to Fig.~\ref{fig:band_trellis} (left).

We now focus on the computation and the storage of the values in the band portion of the trellis.
As $N$ increases, this part dominates the required storage.
We define a new $\hs$-by-$(N-\wi-\wf+3)$ matrix $\boldB$ that represents the non-zero values of $\boldC$ for this band portion.
Accordingly, we denote the column vector that stores the non-zero values in the $n^{\text{th}}$ column of $\boldC$ by $\boldb_n = [b_{\hs,n}\hspace{1.5mm} b_{\hs-1,n}\hspace{1.5mm} \dotsc\hspace{1.5mm} b_{1,n}]^T$ for $n = N-w_{\text{i}}+1, N-w_{\text{i}},\dotsc, w_{\text{f}}-1$.
For the example in Fig.~\ref{fig:band_trellis}, $\boldB$ is given by
\begin{IEEEeqnarray}{rCl} 
    \boldB = [\boldb_2\hspace{0.1cm} \boldb_3\hspace{0.1cm} \boldb_4\hspace{0.1cm} \boldb_5] =
\scalebox{0.9}{$\begin{bmatrix}
    28 & 11 & 4 & 1 \\
    43 & 17 & 7 & 3 \\
    64 & 26 & 10 & 4 \\
    62 & 27 & 12 & 5 \\
    44 & 18 & 8 & 4 \\
    \end{bmatrix}$}.
    \label{eq:stablematrix}
\end{IEEEeqnarray}
Note that we do not start the column indices of $\boldB$ with zero to be consistent with the indices of $\boldC$.
This can be confirmed by comparing \eqref{eq:stablematrix} and Fig.~\ref{fig:band_trellis}.
Note also that $\boldB$ consists of {\it some} columns of $\boldC$ (the band portion) without the zero entries.

Similar to~\eqref{eq:ess_transformation}, $\boldB$ can be computed recursively via
\begin{equation}
    \boldb_{n} = \boldsymbol{A} \boldb_{n+1}, \label{eq:bess_transformation}
\end{equation}
initialized with $\boldb_{N-\wi+1}$ (i.e., the rightmost column of the band portion) and using an $\hs$-by-$\hs$ {\it B-ESS adjacency matrix} $\boldA$.
For the example in Fig.~\ref{fig:band_trellis}, the adjacency matrix is given by
\begin{equation}\tag{10}
\boldA =
\scalebox{0.9}{$
\begin{bmatrix}
1&1&0&0&0 \\
0&1&1&0&0 \\
1&0&1&1&0 \\
0&1&0&1&1 \\
0&0&1&0&1 \\
\end{bmatrix}$}.
\label{eq:bess_adjacency_sone}
\end{equation}
The structure of $\boldA$ depends on $\calA$, $s$, and $\hs$.

Finally, we define $\boldsymbol{r}_n = [r_{\hs,n}\hspace{1.5mm} r_{\hs-1,n}\hspace{1.5mm} \dotsc\hspace{1.5mm} r_{1,n}]^T$, the {\it ratio} vector that stores the element-wise ratio $r_{i,n} = b_{i,n}/b_{i,n+1}$ of adjacent columns $\boldb_{n}$ and $\boldb_{n+1}$ for $ i = 1, 2,\dotsc, \hs$.
As an example, for the trellis in Fig.~\ref{fig:band_trellis} (i.e., for the $\boldB$ in~(\nameref{label:bandtrellismatrix})), $\boldsymbol{r}_3 = [11/4\hspace{1.5mm} 17/7\hspace{1.5mm} 26/10\hspace{1.5mm} 27/12\hspace{1.5mm} 18/8]^T$.

\subsection{Evolution of the Band-trellis}
An interesting property of the band-trellis is that for large enough $N$, elements of $\boldsymbol{r}_n$ become identical as $n$ approaches $w_{\text{f}}$.
In other words, as you move towards node $(0,1)$, i.e., towards the left, adjacent columns become {\it approximately} linked to each other by a scalar multiplicative factor.
Moreover, this scalar multiplicative factor $\rho(\boldA)$ is the spectral radius of $\boldA$, i.e., the greatest (in absolute value) eigenvalue of $\boldA$.
This is a result of the Perron-Frobenius theorem~\cite[Sec. 8.3]{meyer2001_linearalgebra}. 

\begin{figure}[t]
\centering
\includegraphics[width=0.8\columnwidth]{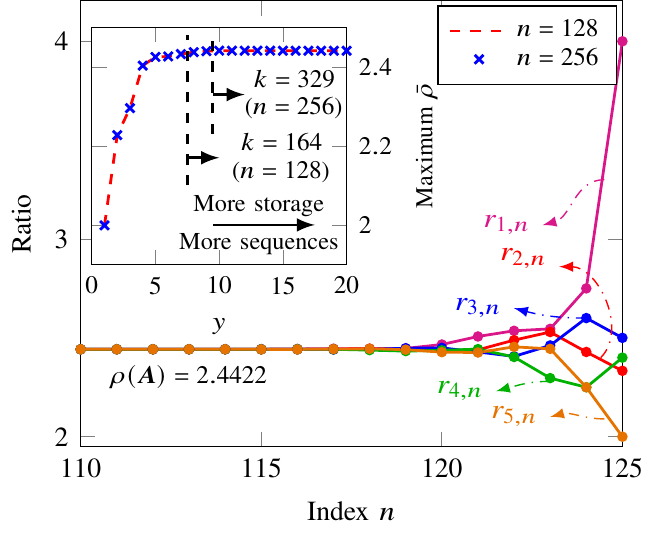}
    \caption{The element-wise ratio of adjacent columns of the band. {\bf Inset:} The maximum $\bar{\rho}$ that satisfies \eqref{eq:ess_summation} vs. the number of columns that are stored at the rightmost part of the band.} 
    \label{fig:ratio_evolve}
\end{figure}

\begin{example}[{\bf Growth of the Band with Rate $\rho(\boldA)$}]\label{ex:scalar}
Consider the band-trellis constructed for $N=128$, $L=129$,  $\calA=\{1, 3, 5, 7\}$, $h_{\text{i}}=w_{\text{i}}=3$, and $s=1$.
The final width of the trellis is $w_{\text{f}} = N+1-(L-h_{\text{i}})/s =3$.
The height of the band is $\hs = h_{\text{i}} + s(w_{\text{i}}-1) = 5$, and thus, the ratio vector $\boldsymbol{r}_n$ is of length 5 for $n = N-w_{\text{i}}, N-w_{\text{i}}-1, \dotsc,w_{\text{f}}-1 = 125, 124, \dotsc, 2$.
The adjacency matrix $\boldA$ for this trellis is already given in \eqref{eq:bess_adjacency_sone}.
The eigenvalues of $\boldA$ are $1.28\exp(\pm77\sqrt{-1})$, $2.4422$, $1$, and $1$.
Thus, $\rho(\boldA)=2.4422$.
In Fig.~\ref{fig:ratio_evolve}, we plotted the elements of $\boldsymbol{r}_n$ as a function of $n$.
We see that the ratios quickly converge to $\rho(\boldA)$.
At $n=110$ and $60$ (not shown in Fig.~\ref{fig:ratio_evolve}), the ratios are equal to $\rho(\boldA)$ up to the fifth and twelfth digit after the decimal point, resp.
\end{example}

Using the convergence of $r_{i,n}$'s, the band portion of the trellis can be {\it approximately} computed via
\begin{equation}\tag{12}
    \boldb_n \approx \rho(\boldA) \boldb_{n+1}, \label{eq:bess_multiplication}
\end{equation}
instead of~\eqref{eq:bess_transformation}, initialized with $\boldb_{N-\wi+1}$.
However, the corresponding $\boldC$ does not necessarily satisfy \eqref{eq:ess_summation} (thus, invertibility is not guaranteed), especially as $n$ approaches $\n$, due to the fluctuation of the ratios from $\rho(\boldA)$, see Fig.~\ref{fig:ratio_evolve}.

To avoid this problem, we propose to store $y$ additional columns $\boldb_{N-\wi-y+1}, \boldb_{N-\wi-y+2}, \dotsc, \boldb_{N-\wi}$ from the band portion (in addition to the initial portion of the trellis and $\boldb_{N-\wi+1}$), and select a $\bar{\rho}<\rho(\boldA)$.
Then we can use 
\begin{equation}\tag{13}
    \boldb_n = \bar{\rho}\boldb_{n+1} \label{eq:bess_multiplication_smallerrho}
\end{equation} 
instead of \eqref{eq:bess_multiplication}, initialized with $\boldb_{N-\wi-y+1}$, to compute the trellis.
This {\it backoff} $\rho(\boldA)\rightarrow \bar{\rho}$ should be adjusted such that the $\boldC$ that corresponds to the $\boldB$ computed with \eqref{eq:bess_multiplication_smallerrho} satisfies~\eqref{eq:ess_summation}.
Though it will inevitably lead to some rate loss, since~\eqref{eq:ess_summation} will not be satisfied with equality.

\begin{example}[{\bf Growth of the Band with Rate $\bar{\rho} < \rho(\boldA)$}]\label{ex:scalar_approx}
 In the inset of Fig.~\ref{fig:ratio_evolve}, we show the maximum $\bar{\rho}$ that satisfies \eqref{eq:ess_summation} when used in \eqref{eq:bess_multiplication_smallerrho} to compute the trellis for the set of parameters given in Example~\ref{ex:scalar}.
 We also show the behavior when $N$ and $L$ are increased to 256 and 257, resp.
 The curves are plotted as a function of the number $y$ of the columns that are stored from the band portion (in addition to the initial portion of the trellis and $\boldb_{N-\wi+1}$).
It is seen that as the number of stored columns $y$ increases, the required backoff $\rho(\boldA)-\bar{\rho}$ decreases.
This is because the ratios $r_{i,n}$ converge to $\rho(\boldA)$ as shown in Fig.~\ref{fig:ratio_evolve}.
On the other hand, for small values of $y$, the backoff is too large that the resulting rate loss decreases the input length of the shaper significantly.
To keep $k=164$ and $329$ for $N=128$ and $N=256$, resp., which are the values that are obtained with the exact computation in \eqref{eq:bess_transformation}, we need to store at least 7 and 9 columns from the band, resp.
\end{example}

\subsection{Trellis Computation via Binary Shifts}\label{ssec:binshift}
We now focus on the practical case in which B-ESS is implemented on hardware using a binary number representation.
Moreover, we assume that the trellis is stored with BP as explained in Sec.~\ref{ssec:bp_ess}: 
Each $c_{l,n}$ is stored with an $n_\alpha$-bit mantissa and an $n_\beta$-bit exponent. 

The approximate relation in \eqref{eq:bess_multiplication} can be extended to 
\begin{equation}\tag{14}
    \boldb_{n} \approx \rho(\boldA)^p \boldb_{n+p}. \label{eq:bess_multiplication_period}
\end{equation}
If here $p$ is selected such that $\rho(\boldA)^p$ is {\it slightly} larger than an integer $t$ power of $2$, we can write \eqref{eq:bess_multiplication_period} as
\begin{equation}\tag{15}
    \boldb_{n} \approx 2^t \boldb_{n+p}, \label{eq:bess_multiplication_binary}
\end{equation}
which can be used to compute the trellis via $t$-bit shifts.
The ``slightly larger'' condition is required to satisfy the invertibility constraint \eqref{eq:ess_summation}.
For a column $\boldb_{n+p}$ of which the elements are stored by mantissa-exponent pairs, \eqref{eq:bess_multiplication_binary} tells that elements of $\boldb_n$ have the same mantissas, and their exponents are increased by $t$.
In such a case, we only need to store the first $y \geq p$ additional columns $\boldb_{N-\wi-y+1}, \boldb_{N-\wi-y+2}, \dotsc, \boldb_{N-\wi}$ from the band portion (in addition to the initial portion and $\boldb_{N-\wi+1}$).
Then the rest can be computed in a block-wise fashion, $p$ columns at a time, by just increasing the exponents of the previous $p$ columns by $t$, i.e., by just appending $t$ zeros to the content of previous $p$ columns.
We call this \emph{shift-based B-ESS}.
Depending on the backoff $\rho(\boldA)^p\rightarrow 2^t$ and the value of $y$, a rate loss will be experienced as discussed in Example~\ref{ex:binary_shifts}.

\subsection{Storage Complexity of B-ESS}
For full precision (FP) ESS, the whole trellis $\boldC$ must be stored with $k$-bit precision which requires a memory of approximately $\n Lk$~bits which scales with $\n^3$~\cite[Sec. IV-B]{GultekinHKW2019_ESSforShortWlessComm}.
For BP ESS, the whole trellis $\boldC$ must be stored with $(n_{\alpha}+n_{\beta})$-bit precision which requires a memory of approximately $\n L(n_{\alpha}+n_{\beta})$~bits which scales with $\n^2\log\n$~\cite[Sec. IV-B]{GultekinHKW2019_ESSforShortWlessComm}.

The number of nodes in a band-trellis can approximately be written as $\wi(\hi+\hs)/2 + \hs(\n-\wf-\wi) + \wf(\hs+1)/2$, where the first, second, and third addends are due to the initial, band, and final portions, resp., see Fig.~\ref{fig:band_trellis}.
Then it can be shown that the memory required to store the values in this many nodes with $(n_{\alpha}+n_{\beta})$-bit precision scales with $\n\log\n$.
For the shift-based B-ESS introduced in Sec.~\ref{ssec:binshift}, the required storage is approximately $(\wi(\hi+\hs)/2 + y\hs)(n_{\alpha}+n_{\beta})$ which scales with $\log\n$ due to the scaling of $n_{\beta}$ in \eqref{eq:approx_ess_exponent}.
By selecting a high enough $n_{\beta}$ (e.g., 16 bits) the storage complexity of shift-based B-ESS can be made independent of $\n$.

\begin{example}[{\bf Growth of the Band with Binary Shifts}]\label{ex:binary_shifts}
Consider the set of parameters given in Example~\ref{ex:scalar} for which $\rho(\boldA) = 2.4422$, and thus, $\rho^7 = 518.24 > 2^9$, i.e., $p=7$ and $t = 9$.
Here we switch to the BP implementation, i.e., instead of storing the numbers with FP using $k+1=165$~bits, we only use $n_\alpha+n_\beta = 10+8=18$~bits.
With this approximation, $k$ is still $164$ (i.e., no rate loss due to BP).
Then, by only storing 14 columns ($\wi-1=2$ from the initial portion, $y+1=12$ from the band), the rest of the trellis can be computed by appending $p=7$ new columns recursively via \eqref{eq:bess_multiplication_binary} as $\boldb_n \approx 2^9\boldb_{n+7}$.
In the mantissa-exponent representation, this means the mantissas of $b_{i,n}$ and $b_{i,n+7}$ are the same, while the former has its exponent increased by 9.
When $\boldB$ is computed with this approximation, $k$ is still $164$ for the corresponding $\boldC$ (i.e., no rate loss due to shift-based computation~\eqref{eq:bess_multiplication_binary}). 
Therefore, although the number of sequences decreased due to BP and due to the approximation in~\eqref{eq:bess_multiplication_binary}, the rate $k/\n$ stayed the same.
However, the required storage which was $13.01$ and $1.42$ kB in the FP and BP cases, resp., is now 0.15 kB. 
\end{example}

\section{Numerical Results and Discussion}

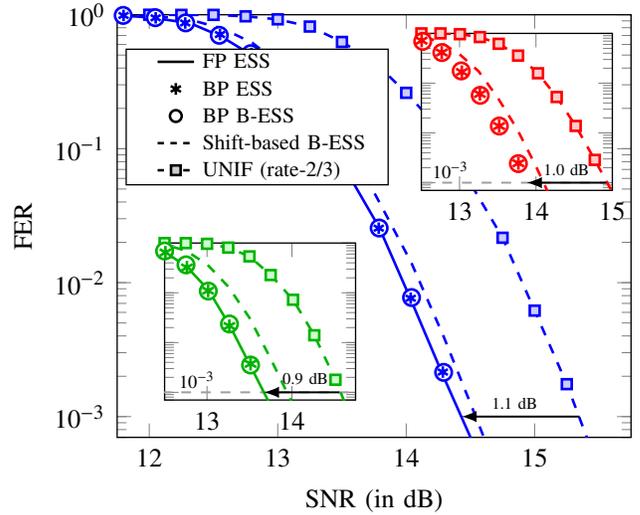
\begin{figure}[t]
    \centering
    \resizebox{0.95\columnwidth}{!}{\begin{tikzpicture}
\usepgfplotslibrary{fillbetween}
\begin{axis}[
name=outset,
  every axis/.append style={font=\footnotesize},
width=0.8\columnwidth,
xmin=11.75,
xmax=15.75,
ymin=7e-4,
ymax=1,
grid style={dashed,lightgray!75},
ymode = log,
xlabel={SNR (in dB)},
ylabel={FER},
ylabel near ticks,
xlabel near ticks,
legend style={at={(0.2,0.2)},anchor=center,font=\footnotesize,legend cell align=left,row sep=-0.75ex},
]
\addplot [color=black, very thick, dashed, mark=square*, mark size=2.25pt,mark options={fill=black!20!white,solid}]
  table[row sep=crcr]{
1 1\\
};
\label{line:unif} 
\addplot [color=black, very thick, no marks]
  table[row sep=crcr]{
1 1\\
};
\label{line:fp}
\addplot [color=black, very thick, only marks, mark=asterisk, mark size=2.75pt]
  table[row sep=crcr]{
1 1\\
};
\label{line:bp}
\addplot [color=black, very thick, dashed, mark=*, only marks, mark size=2.75pt,mark options={fill=white!20!white,solid}]
  table[row sep=crcr]{
1 1\\
};
\label{line:bessfull}
\addplot [color=black, very thick, dashed]
  table[row sep=crcr]{
1 1\\
};
\label{line:bessshiftclose}
\addplot [color=blue, thick, dashed, mark=square*, mark size=1.5pt,mark options={fill=blue!20!white,solid}]
  table[row sep=crcr]{
12	1\\
12.2500000000000	0.997524752475248\\
12.5000000000000	0.997524752475248\\
12.7500000000000	0.973253379021512\\
13	0.924083561698241\\
13.2500000000000	0.816373141146018\\
13.5000000000000	0.626605596954671\\
13.7500000000000	0.446935506278545\\
14	0.261250682091163\\
14.2500000000000	0.143400437963308\\
14.5000000000000	0.0551617605994673\\
14.7500000000000	0.0216361584331294\\
15	0.00618423620515965\\
15.2500000000000	0.00174673877063229\\
15.5000000000000	0.000402500000000000\\
15.7500000000000	0.000107500000000000\\
};
\addplot [color=blue, thick, no marks]
  table[row sep=crcr]{
11.0404936533276	1\\
11.2865335845735	1\\
11.5417192703978	1\\
11.7762892833877	0.987816485817628\\
12.0159983204158	0.952596978403415\\
12.3201050119213	0.863124175289951\\
12.5568311517954	0.691117024216193\\
12.7991453412656	0.498220095931075\\
13.0515688462989	0.306126201183376\\
13.2772517648210	0.175999887676752\\
13.5456385586014	0.0654363096100284\\
13.7903849072460	0.0257373440169135\\
14.0417104914380	0.00725156689440065\\
14.2904110654354	0.00198620493498958\\
14.5409205858242	0.000572500000000000\\
}; 
\addplot [color=blue, thick, mark=*, only marks, mark size=2.5pt,mark options={fill=white,solid}]
  table[row sep=crcr]{
11.0356374960541	1\\
11.2885146968229	0.999335989375830\\
11.5509689374189	0.997349227428909\\
11.7968891671414	0.988510198703184\\
12.0471505830002	0.950452168484225\\
12.2834772227987	0.869569139251239\\
12.5451735512618	0.702407007334830\\
12.7894745300067	0.512266430699053\\
13.0424185568250	0.310253092632152\\
13.2932704657236	0.161002018526150\\
13.5432475290887	0.0720416331513502\\
13.7896831265309	0.0255642386035365\\
14.0414389593533	0.00774173568894253\\
14.2912713606185	0.00213865407659296\\
14.5408590302549	0.000507274915773949\\
};
\addplot [color=blue, thick, only marks, mark=asterisk, mark size=1.75pt,mark options={fill=blue,solid}]
  table[row sep=crcr]{
11.0301216490490	1\\
11.2855985122920	1\\
11.5383127547673	1\\
11.8	0.987816485817628\\
12.05	0.948319618605490\\
12.28	0.868017496561069\\
12.5503612277050	0.711543903202228\\
12.7944952463062	0.513589910674616\\
13.0432194190387	0.311219730405127\\
13.2692227478196	0.180019794487816\\
13.5412340634178	0.0755\\
13.7861891476348	0.0255\\
14.0350119115517	0.00765\\
14.2851412343510	0.00215\\
14.5356578638853	0.000477500000000000\\
};
\addplot [color=blue, thick, dashed, no marks]
  table[row sep=crcr]{
10.5368986518824	1\\
10.7954811816359	1\\
11.0433402533332	1\\
11.2719692826543	1\\
11.5164538486600	1\\
11.8257424463584	0.992574257425743\\
12.0606280851957	0.968986098105960\\
12.2878656468410	0.934703060871285\\
12.5687862778896	0.789404227176814\\
12.8000095958835	0.619583399479066\\
13.0344183180163	0.413914356850390\\
13.2651839058350	0.254420908282860\\
13.5464933037943	0.108161304818663\\
13.7933204515216	0.0394850076187629\\
14.0397898098652	0.0140851058424680\\
14.2921672317545	0.00394248528445925\\
14.5404475159938	0.00100366887849297\\
14.7905755446350	0.000244333333333333\\
};
\draw[-latex,semithick] (15.35,1e-3) -- node[above, inner sep=2pt, pos=0.55] {\tiny 1.1 dB} (14.43,1e-3);
\legend{};
\end{axis}
\begin{axis}[name=insetfig, 
draw, width=0.2\textwidth,
height=0.18\textwidth, 
every axis/.append style={font=\footnotesize},
anchor=south east, at={(outset.south east)},
yshift=0.4cm,
xshift=-2.95cm,
axis background/.style={fill=white},
xmin=12.5,
xmax=14.75,
ymin=7e-4,
ymax=1,
ymode = log,
ylabel near ticks,
xlabel near ticks,
 yticklabels={,,},
legend style={at={(0.24,0.15)},anchor=center,font=\tiny,legend cell align=left,row sep=-1ex},
]
\draw[-, dashed, semithick, gray!75] (12,1e-3)--(16,1e-3) node[black,above, pos=0.25] {};
\draw[draw=none, dashed, semithick, gray!75] (12,7e-4)--(16,7e-4) node[black,above, pos=0.22] {\scalebox{0.7}{$10^{-3}$}};
\addplot [color=green!70!black, thick, dashed, mark=square*, mark size=1.5pt,mark options={fill=green!20!white,solid}]
  table[row sep=crcr]{
  12	1\\
12.2500000000000	1\\
12.5000000000000	1\\
12.7500000000000	0.997524752475248\\
13	0.964187493776201\\
13.2500000000000	0.809776116519381\\
13.5000000000000	0.539627823722349\\
13.7500000000000	0.227570703656825\\
14.0134513491768	0.0724407893352840\\
14.2638861710574	0.0140093520291724\\
14.5130082520857	0.00178920671522175\\
14.7636204917344	0.000161000000000000\\
};
\addplot [color=green!70!black, thick, no marks]
  table[row sep=crcr]{
10.5423786584476	1\\
10.7665478626999	1\\
11.0223464889341	1\\
11.2687589485487	1\\
11.4875386597602	1\\
11.7427888601671	1\\
12.0237545778092	0.985341238292876\\
12.2451964985593	0.914121505935665\\
12.5226820521686	0.629825382063234\\
12.7580669083954	0.336914552886749\\
13.0119889354874	0.116363272180767\\
13.2646063353881	0.0223921480277857\\
13.5133071464160	0.00334028534661972\\
13.7634819290192	0.000485000000000000\\
14.0135000177917	7.50000000000000e-05\\
};
\addplot [color=green!70!black, thick, mark=*, only marks, mark size=2.5pt,mark options={fill=white,solid}]
  table[row sep=crcr]{
11.5406898454138	1\\
11.7659408900045	1\\
12.0220344144924	0.985341238292876\\
12.2692667421217	0.882016268045680\\
12.5	0.691470631619895\\
12.7522087591614	0.364885965643939\\
13.0147641278728	0.108314776633586\\
13.2643810788513	0.0236280010184057\\
13.5124308517629	0.00353614756555381\\
13.7626067762057	0.000457500000000000\\
14.0127821615675	8.00000000000000e-05\\
}; 
\addplot [color=green!70!black, thick, only marks, mark=asterisk, mark size=2pt,mark options={fill=green!70!black,solid}]
  table[row sep=crcr]{
10.4874492530261	1\\
10.7569319298876	1\\
10.9888903156939	1\\
11.2748490392454	1\\
11.4984648311188	1\\
11.7694363895556	1\\
12.0230732710581	0.978242782261420\\
12.2511282182541	0.928128999272787\\
12.5099223023409	0.645717048608309\\
12.7632652848238	0.323074926859438\\
13.0127985517277	0.110242438535824\\
13.2628778200943	0.0215877478265020\\
13.5128142609242	0.00372748201994127\\
13.7626267534625	0.000472500000000000\\
14.0140695101207	6.25000000000000e-05\\
};
\addplot [color=green!70!black, thick, dashed, no marks]
  table[row sep=crcr]{
10.7500000000000	1\\
11	1\\
11.2500000000000	1\\
11.5000000000000	1\\
11.7500000000000	1\\
12	1\\
12.2500000000000	0.989490560593722\\
12.5000000000000	0.928277128796868\\
12.7500000000000	0.703736897420386\\
13	0.374471733348310\\
13.2500000000000	0.133130583550854\\
13.5000000000000	0.0314125669518287\\
13.7500000000000	0.00472220781802129\\
14	0.000624153468708508\\
14.2500000000000	9.51666666666667e-05\\
};
\draw[-latex,semithick] (14.56,1e-3) -- node[above, inner sep=2pt, pos=0.45] {\tiny 0.9 dB} (13.67,1e-3);
\legend{};
\end{axis}
\begin{axis}[name=insetfig2, 
draw, width=0.2\textwidth,
height=0.18\textwidth, 
every axis/.append style={font=\footnotesize},
anchor=north east, at={(outset.north east)},
yshift=-0.2cm,
xshift=-0.2cm,
axis background/.style={fill=white},
xmin=12.5,
xmax=15,
ymin=7e-4,
ymax=1,
ymode = log,
ylabel near ticks,
xlabel near ticks,
 yticklabels={,,},
legend style={at={(0.24,0.15)},anchor=center,font=\tiny,legend cell align=left,row sep=-1ex},
]
\draw[-, dashed, semithick, gray!75] (12,1e-3)--(16,1e-3) node[black,above, pos=0.25] {};
\draw[draw=none, dashed, semithick, gray!75] (12,7e-4)--(16,7e-4) node[black,above, pos=0.22] {\scalebox{0.7}{$10^{-3}$}};
\addplot [color=red, thick, dashed, mark=square*, mark size=1.5pt,mark options={fill=red!20!white,solid}]
  table[row sep=crcr]{
  12.5 1\\
  12.75 1\\
13.0037962491792	0.964507899147705\\
13.2662551007314	0.844242332379173\\
13.5127133291905	0.618190832307781\\
13.7663580166778	0.357958026318891\\
14.0270249092710	0.158014110248818\\
14.2718166475802	0.0527353014591807\\
14.5202870307301	0.0138345389196643\\
14.7699697862985	0.00287033038415761\\
15.0208523540770	0.000562500000000000\\
};
\addplot [color=red, thick, no marks]
  table[row sep=crcr]{
10.7666014353525	1\\\\
11.0126546971261	1\\\\
11.2769465900185	1\\\\
11.5244052348188	1\\\\
11.7842633209134	0.995098039215686\\\\
12.0181421494713	0.975911723912866\\\\
12.2689348521323	0.883505281945372\\\\
12.5407903947502	0.676840208240016\\\\
12.7719111918971	0.402787920463823\\\\
13.0164206792978	0.181897313971669\\\\
13.2728044578452	0.0549452876658473\\\\
13.5221256549235	0.0124586165922865\\\\
13.7697130197721	0.00238299853530013\\\\
14.0202401657532	0.000310000000000000\\\\
14.2702987106725	6.00000000000000e-05\\\\
};
\addplot [color=red, thick, mark=*, only marks, mark size=2.5pt,mark options={fill=white,solid}]
  table[row sep=crcr]{
12.5175140843479	0.691360982167678\\
12.7685401402886	0.411625935618001\\
13.0214502419269	0.176248221846047\\
13.2706064895363	0.0569653905065953\\
13.5204658692989	0.0137494095673728\\
13.7698690820547	0.00241933605415825\\
};
\addplot [color=red, thick, only marks, mark=asterisk, mark size=2pt,mark options={fill=red,solid}]
  table[row sep=crcr]{
10.7523844385454	1\\
11.0372602484095	1\\
11.2761517011148	1\\
11.5093834473786	1\\
11.7780997966301	0.995098039215686\\
12.0095657661762	0.971056883774369\\
12.2795853238138	0.881759388611931\\
12.5033673671007	0.714886026752153\\
12.7667877655669	0.411502680257752\\
13.0295825447841	0.168010585519352\\
13.2660621609795	0.0604635820325126\\
13.5188281162331	0.0137386300317773\\
13.7702061383590	0.00245921098932639\\
14.0193311898547	0.000287500000000000\\
14.2698399359376	5.00000000000000e-05\\
};
\addplot [color=red, thick, dashed, no marks]
  table[row sep=crcr]{
11	1\\
11.2500000000000	1\\
11.5000000000000	1\\
11.7500000000000	0.999667994687915\\
12	0.997351831147085\\
12.2500000000000	0.977562726141581\\
12.5000000000000	0.890296467858910\\
12.7500000000000	0.663151586876313\\
13	0.381467867118637\\
13.2500000000000	0.172651022807921\\
13.5000000000000	0.0509020708319157\\
13.7500000000000	0.0122181820604821\\
14	0.00213802022268177\\
14.2500000000000	0.000317500000000000\\
14.5000000000000	5.48333333333333e-05\\
};
\draw[-latex,semithick] (14.92,1e-3) -- node[above, inner sep=2pt, pos=0.5] {\tiny 1.0 dB} (13.88,1e-3);
\legend{};
\end{axis}
\node[draw, fill=white, inner sep=0pt, anchor = south west] at (0.09,2.7) {
\addtolength{\tabcolsep}{-4.25pt}  
\scalebox{0.65}{
\renewcommand{\arraystretch}{1.06}
\begin{tabular}{cl}
    \ref{line:fp} & FP ESS \\
    \ref{line:bp} & BP ESS \\
    \ref{line:bessfull} &  BP B-ESS \\
    \ref{line:bessshiftclose} &  Shift-based B-ESS \\
    \ref{line:unif} & UNIF (rate-2/3) \\
\end{tabular}
}
\addtolength{\tabcolsep}{4.25pt}
};
\end{tikzpicture}}
    \caption{Frame error rates (FERs) vs. SNR with 648- (blue, main), 1296- (red, top-right inset), and 1944-bit (green, bottom-left inset) FEC codes.}
    \label{fig:FERvsSNR}
\end{figure}

Fig.~\ref{fig:FERvsSNR} shows the performance of PAS (see Fig.~\ref{fig:pas}) combining different ESS algorithms of shaping rate $k/\n=1.5$ b/amp at $\n=216$, $432$, and $648$ with the IEEE 802.11's 648-bit (blue), 1296-bit (red), and 1944-bit (green), resp., low-density parity-check codes of rate $5/6$.
With 8-ASK, this combination gives a transmission rate of $4$ b/2D, see~\cite[Table IV]{GultekinHKW2019_ESSforShortWlessComm}, which can be obtained with rate-$2/3$-coded uniform signaling, also shown in the figure.
The corresponding storage complexities are highlighted with the same color code in Table~\ref{tab:compare_rates}.
FP ESS provides around 1 dB gain over uniform signaling, though it has storage complexity on the order of MBs, scaling with $\n^3$.
BP ESS achieves virtually the same performance with a storage requirement below an MB, scaling with $\n^2\log\n$.
However, BP B-ESS ($\boldC$ stored completely, see the third section in Table~\ref{tab:compare_rates}) provides again the same performance with around a 75\% reduction in required storage which scales with $\n\log\n$.
The band-trellis has a band width of $\hs\geq55$ demonstrating that B-ESS approaches ESS as $\hs$ increases. 
However, ESS-level performance can be achieved without having $\hs=L$, which is the reason behind the decrease in required storage.

\begin{table}[t]
\caption{Scaling of the Storage Complexity of ESS and B-ESS ($s=1$)\label{tab:compare_rates}}
\renewcommand{\arraystretch}{1.4}
\centering
\resizebox{\columnwidth}{!}{
\begin{tabular}{|c|c|c|c|c|c|l|} 
\hline 
& Fig.~\ref{fig:FERvsSNR} & $y$ & $N$ & $L$ & $\hi, \wi, \rho, \hs, t, p$  & Storage ($n_\alpha, n_\beta$) \\
\hline
\multirow{3}{*}{\rotatebox[origin=c]{90}{FP ESS}} & \multirow{3}{*}{\rotatebox[origin=c]{90}{(\ref{line:fp})}} & & 216 & 184 & $\hi=L$ & \cellcolor{blue!15}1.61 MB \\
& & $L$ & 432 & 361 & $\wi=\n+1$  & \cellcolor{red!20}12.63 MB \\
& &  & 648 & 538 & $\hs=L$ & \cellcolor{green!20}42.36 MB \\
\hline
\multirow{3}{*}{\rotatebox[origin=c]{90}{BP ESS}} & \multirow{3}{*}{\rotatebox[origin=c]{90}{(\scalebox{1}{\ref{line:bp}})}}& & 216 & 184 & $\hi=L$ & \cellcolor{blue!15}94.39 kB $(10, 9)$ \\
& & $L$ & 432 & 361 & $\wi=\n+1$  & \cellcolor{red!20}409.37 kB $(11, 10)$ \\
& &  & 648 & 538 & $\hs=L$ & \cellcolor{green!20}958.72 kB $(12, 10)$ \\
\hline
\multirow{3}{*}{\rotatebox[origin=c]{90}{B-ESS}} & \multirow{3}{*}{\rotatebox[origin=c]{90}{(\ref{line:bessfull})}} &  & 216 & 184 & $\hi=16, \wi=40$  & \cellcolor{blue!15}23.22 kB ($10, 9$)\\
 & & $L$ & 432 & 361 & $\hi=16, \wi=84$ & \cellcolor{red!20}104.91 kB ($14, 10$)\\
  & &  & 648 & 538 & $\hi=16, \wi=124$ &  \cellcolor{green!20}221.05 kB ($14, 10$) \\
\hline
\multirow{3}{*}{\rotatebox[origin=c]{90}{Shift-based}} & \multirow{3}{*}{\rotatebox[origin=c]{90}{(\ref{line:bessshiftclose})}} &  & 216 & 190 & $\hi=16, \wi=40$   & \cellcolor{blue!15}4.51 kB ($13, 9$)\\
  & & 4  & 432 & 383 & $\rho(\boldA)=3.0662$ & \cellcolor{red!20}4.51 kB ($13, 9$)\\
  & &    & 648 & 579 & $\hs=55, t=8, p=5$      & \cellcolor{green!20}4.51 kB ($13, 9$) \\
\hline
\end{tabular}
}
\end{table}

Fig.~\ref{fig:FERvsSNR} also shows the performance of B-ESS implemented via the shift-based technique explained in Sec.~\ref{ssec:binshift}. 
We see that with this technique, it is possible to operate within less than 0.2 dB of ESS when a relatively wide band width of $\hs=55$ is considered (see the fourth section in Table~\ref{tab:compare_rates}).
In return, a significant decrease in required storage is achieved concerning B-ESS with pre-computed-and-stored $\boldC$: $4.5$ kB is required instead of $23.2$, $104.9$, and $221.1$ kB for $\n=216$, $432$, and $648$, resp.
This storage reduction is thanks to the shift-based computation that allows us to store only $y+wi=4+40=44$ columns of $\boldC$.
The minor loss in the performance is also due to the shift-based computation which results in rate loss due to the backoff $\rho^p \rightarrow 2^t$ (see~\eqref{eq:bess_multiplication_period} and~\eqref{eq:bess_multiplication_binary}).

Finally, we emphasize that since the shift-based B-ESS requires only the storage of a small number of columns, for fixed $\hi$ and $\wi$, $\n$ can be increased arbitrarily without incurring any increase in storage complexity.
This can be seen from the constant required storage in the fourth section of Table~\ref{tab:compare_rates}.
This paves the way for a streaming amplitude shaper, operating similar to a convolutional encoder, in which the trellis is computed in a time-invariant manner, i.e., via binary shifts.
This can change the current amplitude shaping paradigm which typically assumes a fixed $\n$ and block-wise processing. 
Moreover, since B-ESS generalizes ESS as we discussed above, this streaming mode operation can find application in a variety of communication scenarios ranging from the linear AWGN channel to nonlinear fiber channels.

\section{Future Outlook}
Performance evaluation of shift-based B-ESS for the nonlinear fiber channel (for which narrow band-trellises are convenient to produce sequences with small energy variations~\cite{gultekin_bess_ofc_arxiv}) is left for future work.
Also, streaming shapers operating similarly to convolutional encoders immediately call for Viterbi algorithm-based deshapers. 
Thus, future research avenues include Viterbi (soft) deshaping, iterative soft information exchange between the decoder and the deshaper, and joint decoding/deshaping on the product of shaping and coding trellises.
These were hitherto deemed impractical due to the large size of ESS trellises, but now seem viable with B-ESS.

\begin{spacing}{0.8}
{\small {\bf Acknowledgements:} The work of Y.C. G\"{u}ltekin and A. Alvarado has received funding from the ERC under the EU's H2020 programme via the Starting grant FUN-NOTCH (ID: 757791) and via the Proof of Concept grant SHY-FEC (ID: 963945).}
\end{spacing}

\bibliographystyle{IEEEtran}
\bibliography{IEEEabrv,PhD_refs}

\end{document}